\documentclass[12pt]{article}

\usepackage{bbold}
\usepackage{amssymb}
\usepackage{slashed}
\usepackage{graphics}
\usepackage{epsfig}

\declareslashed{}{/}{.1}{0}{E}

\def\l{\lambda}

\def\wh#1{\widehat{#1}}
\def\wt#1{\widetilde{#1}}

\def\d{\partial}

\def\be{\begin{equation}}
\def\ee{\end{equation}}
\def\beq{\begin{equation}}
\def\eeq{\end{equation}}
\def\bea{\begin{eqnarray}}
\def\eea{\end{eqnarray}} 
\def\beqa{\begin{equation}\begin{array}{l}}
\def\eeqa{\end{array}\end{equation}}

\def\eqn#1{(\ref{#1})}

\def\eqref#1{eq.~(\ref{eq:#1})}

\def\boxfrac#1#2{\mbox{\small{$\frac{#1}{#2}$}}}
\def\half{\mbox{\small{$\frac{1}{2}$}}}

\def\quarter{\mbox{\small{$\frac{1}{4}$}}}

\def\nn{\nonumber}

\begin{document}

\vspace{.8cm}
\setcounter{footnote}{0}
\begin{center}
{\Large{\bf 
Higher Spin Gravitational Couplings and the\\[4mm] Yang--Mills Detour Complex
    }}\\[10mm]

{\sc \small A.R.\ Gover\\[4mm]}

{\em\small  
Department of Mathematics, The University of Auckland, New Zealand\\ 
{\tt gover@math.auckland.ac.nz}
}\\[7mm]

{\sc \small K.\  Hallowell and A.\ Waldron\\[4mm]}

{\em\small  
Department of Mathematics, University of California, Davis,
CA 95616, USA\\ 
{\tt hallowell,wally@math.ucdavis.edu}
}\\[5mm]

\bigskip

\bigskip

{\sc Abstract}\\

\end{center}

{\small
\begin{quote}

Gravitational interactions of higher spin fields are generically plagued by inconsistencies.
We present a simple framework that couples 
higher spins to a broad class of gravitational backgrounds (including Ricci flat and Einstein)
consistently at the classical level. The model is the simplest example of
a Yang--Mills detour complex, which recently has been applied in the mathematical 
setting of conformal geometry.
An analysis of asymptotic scattering states about the
trivial field theory vacuum in the simplest version of the theory yields 
a rich spectrum marred by negative norm excitations.
The result is  a theory of a physical massless graviton, scalar field,
and massive vector along with a degenerate pair of zero norm photon excitations.
Coherent states of the unstable sector of the model do have positive norms,
but their evolution is no longer unitary and their amplitudes grow with time.
The model is of considerable interest  for braneworld scenarios and ghost condensation models, and invariant theory.
 
\bigskip

\bigskip

\end{quote}
}

\newpage



\section{Introduction}

Massless, massive, and partially massless free higher spin fields
propagate consistently in maximally symmetric backgrounds ({\it i.e.},
Minkowski, de Sitter and Anti de Sitter
spaces)~\cite{Singh:1974qz,Vasiliev:1986td,Deser:1983tm}.  Allowing
generic curved backgrounds introduces various
inconsistencies. Firstly, introducing general curvatures
$R_{\mu\nu}{}^\#=[D_\mu,D_\nu]$ can destroy the gauge invariances or
constraints which ensured the correct physical degree of freedom count
of maximally symmetric
backgrounds~\cite{Aragone:1979bm,Buchbinder:1999ar}.  Secondly, even
in benign backgrounds ensuring correct degrees of freedom, signals may
propagate at superluminal speeds~\cite{Velo:1969bt,Deser:2001dt} In
this Article we display a simple mechanism for maintaining the gauge
invariances of higher spins in a broad class of gravitational
backgrounds.

Much mathematical insight into the structure of manifolds has been
gained by studying the equations of mathematical physics. Notable
examples include the self-dual Yang--Mills equations and
Donaldson's four manifold theory, and ensuing simplifications based on
the monopole equations of its
supersymmetrization~\cite{Donaldson:1990kn}. In self-dual Yang-Mills
theory an important {\it r\^ole} is played by a class of two operator
complexes that are sometimes termed Yang-Mills complexes. In
\cite{GSS} it is observed that there is a closely related 3 operator
complex for each full Yang-Mills connection. These are there termed 
Yang-Mills detour complexes since there are intimate links with
conformal geometry and in dimension four the complexes fall into a class of 
complexes called conformal detour complexes \cite{Br-BrGodeRham}.
The Yang-Mills detour complexes are related to an idea that has been
extant in the Physics literature for some time.  Namely, it is well
known that massless vectors couple consistently to an onshell
Yang--Mills background if a non-minimal coupling is
included~\cite{Deser:1987uk}.  Unwrapping this in mathematical terms
yields a Yang-Mills detour complex.  Here we propose to study a
Yang-Mills detour complex in one of the simplest possible settings in
order to expose and explore, for a physics audience, the issues of
consistency at both the classical and quantum level. 
On a dimension~4 Lorentzian background we  obtain a
theory of higher spins  by taking the Poincar\'e group as
Yang--Mills gauge group and the vectors transforming in any finite
dimensional representation.

The first objection, that this simple model mixes spacetime and
internal symmetries, and so violates the Coleman--Mandula
theorem~\cite{Coleman:1967ad}, is evaded because we propose only a
theory of non-interacting free fields whereas the theorem pertains to
triviality of an interacting $S$-matrix. The second complaint that
finite dimensional representations of the non-compact Poincar\'e group
are non-unitary and therefore imply the likelihood of ghost states is,
however borne out.  (We note that an infinite
dimensional unitary representation ought yield an infinite tower of
consistent higher spin interactions and comment further in the
Conclusions.)  In the trivial field theory vacuum we indeed find a
pair of degenerate, zero norm photons. Nonetheless, the model is of
considerable interest because
\begin{enumerate}
\item Ghost states can simply indicate instability of the
trivial Lorentz invariant vacuum. The model is useful as both a laboratory to study these excitations
plus there exists the possibility of finding a (possibly non-Lorentz invariant) stable
vacuum (especially if interactions were included).
\item The model can be used to study properties of the background manifold in which the
higher spin fields propagate. 
Higher spin gauge invariances can provide new invariants of the background manifold~\cite{Gover}.
 Moreover, finding physical states
 amounts to  computing the cohomology of the twisted Maxwell complex. 
\item Backgrounds other than the simplest Minkowski one, may permit a physical scattering
spectrum.
\end{enumerate}

For the simplest non-trivial spin~2 example in a four dimensional
Minkowski background we find the following spectrum\footnote{For flat backgrounds,
the mass parameter $m$ is freely tunable (save to vanishing values). In  general
spaces it depends on the gravitational coupling. A parameter space
study as in~\cite{Deser:2001dt}  is then required.}:

\begin{center}
\begin{tabular}{c|c|c}
Spin & Mass & Norm\\
\hline \hline
2 & 0 & +ve\\
1 & $\sqrt 2 m$ & +ve\\
1 & 0 & 0\\
1 & 0 & 0\\
0 & 0 & +ve\\
\hline
\end{tabular}
\end{center}

The Lorentz invariant Lagrangian for these excitations depends on
(i) a 2-index symmetric tensor, (ii) a 2-form, and  (iii) a vector field.
However a detailed Hamiltonian helicity analysis is required to determine
the graviton, massive vector, two photon, and massless scalar spectrum quoted above.
Interestingly, the photon states correspond to generalized eigenvector solutions
of the wave equations of motion. Physically this amounts to resonance states
with amplitudes growing linearly in time. Moreover, in the unstable photon subspace of
the Hilbert space, only zero norm states diagonalize the Hamiltonian.
Coherent states of these photon excitations have norms which grow with time, in violation of unitarity, 
and signify the instability of the model.

This Article is arranged as follows. In Section~\ref{twist} we explain how to formulate higher
spins as a complex and present the twisted Maxwell complex. In Section~\ref{Minkowski}
we specialize the underlying vector matter fields to the fundamental representation
of the Poincar\'e Yang--Mills gauge group. The Hamiltonian analysis of this model is given
in Section~4 while Section~\ref{Danger} concentrates on the dangerous helicity one
excitations. The quantization of the model is
given in Section~6. In Section~\ref{coherent_state} we compute coherent states
and their evolution.
Our conclusions and further speculations are given in Section~\ref{Conclusions}.

\section{Yang--Mills Detour Complex}

\label{twist}

An obvious, yet powerful, observation is that in any dimension  we can
view a classically consistent higher spin gauge theory as a complex \be 0
\longrightarrow \left\{\!\!\!\!
\begin{array}{cc}
\mbox{\small Gauge}\\[-1mm]
\mbox{\small Parameters}
\end{array}
\!\!\!\!\right\}
\stackrel{\textstyle \cal D}{\longrightarrow} 
\left\{
\mbox{Fields}
\right\}
\stackrel{\textstyle \cal G}{\longrightarrow} 
\left\{\!\!\!\!
\begin{array}{cc}
\mbox{\small Field}\\[-1mm]
\mbox{\small Equations}
\end{array}
\!\!\!\!\right\}
\stackrel{\textstyle {}^*\cal D}{\longrightarrow} 
\left\{\!\!\!\!
\begin{array}{cc}
\mbox{\small Bianchi}\\[-1mm]
\mbox{\small Identities}
\end{array}
\!\!\!\!\right\}
\longrightarrow 0\, .
\label{Maxwell}
\ee Here where we write ``Field Equations'' is really of course the
vector bundle where these equations take values and a similar comment
applies to the ``Bianchi Identities'' which give the integrability
condition for the field equations. 
  The simplest example is the
Maxwell (detour) complex where the space of fields are one forms $V\in \Gamma
(\Lambda^1M)$, and ${\cal D}=d$ the Poincar\'e differential, its dual is
${}^*{\cal D}=*d*$ and Maxwell's equations are simply \be {\cal G} V
\equiv \delta d V = 0\, .  \ee
In this case the statement that~\eqn{Maxwell} is a
complex so that ${\cal G}\, {\cal D} = 0 = {}^*{\cal D} \, {\cal G}$
amounts to the gauge invariance $V\rightarrow V+d\alpha$ and the
Bianchi identity $\delta \, {\cal G} V=0$.

The Maxwell complex can be twisted by coupling to a vector bundle
connection over the manifold $M$. In general then \eqn{Maxwell} fails
to be a complex reflecting the usual problem of adding curvature to a
flat theory. However, if the connection satisfies the Yang-Mills
equations then remarkably it turns out that we still obtain a 
complex called a Yang--Mills detour complex~\cite{GSS} .  Let us review, in our current notation and on a spacetime
background, this simple construction.  In this setting, the space of
fields are one-forms taking values in a representation $R$ of the
Yang-Mills gauge group~$G$.  We work locally, so for the purposes of
the calculations the manifold may be taken to be $\mathbb{R}^4$ and
the bundle carrying the representation may be taken trivial (as a
vector bundle).  Let \be D=d+A\, , \ee be the Yang--Mills connection
(so the Yang--Mills potential $A$ is a ${\mathfrak g}$-valued one-form).  Then we set \bea {\cal D}&=&D\, ,\nn\\ {}^*{\cal D}&=&*\, D*\,
,\nn\\ {\cal G}&=&*\, D*D-*({}^*F)\, , \eea where $F=D^2$ is the
Yang--Mills curvature.  Now we find that~\eqn{Maxwell} is a complex so
long as the Yang--Mills connection obeys the Yang--Mills equations \be
[D,{}^*\!F]=0\, .  \ee Physicists would summarize this information in
terms of the action (valid in any spacetime dimension and signature)
\be S=\frac12 \int_M V_\mu^T \Big(g^{\mu\nu}D^\rho
D_\rho - D^\nu D^\mu + F^{\mu\nu}\Big) V_\nu\, ,
\label{Indices}
\ee
with gauge invariance $V_\mu\rightarrow V_\mu+D_\mu \alpha$ valid whenever $D^\mu F_{\mu\nu}=0$ (suppressing indices corresponding to the representation $R$).

The existence and origin of this model is also clear from a physical standpoint.
Yang-Mills theory itself can be constructed iteratively by coupling vectors to vectors~\cite{Deser:1987uk}. The first
step of coupling abelian vector fields $V$ to the non-abelian vector field $A$
 requires that the field $A$ is on-shell.
 
The model~\eqn{Indices} is a consistent one for any compact gauge group~$G$ and
unitary representation $R$. Our proposal is simply to relax compactness of~$G$ and
take it to be the spacetime Poincar\'e symmetry algebra, and to begin our study with
finite dimensional representations $R$. The ghost difficulties that the model faces are all hidden
in the superscript ``$T$'' on the field $V_\mu$ in~\eqn{Indices}, indicating an inner product on 
vectors in the representation space $R$. 

Nonetheless, the proposal is rather fruitful since taking the gauge group~$G$ to be the Poincar\'e
one amounts to coupling the model to gravity. This idea is well known both in mathematics
and physics (called the Cartan connection or Palatini formalism, respectively).
Let us concentrate on four dimensions and adopt the
5$\times$5 matrix representation of the Poincar\'e Lie algebra so that the background Yang--Mills
potential reads
\bea
A = \left (
\begin{array}{cc}
\omega^m{}_n & e^m \\
0 & 0
\end{array}
\right ),
\eea
where indices $m,n,..$ take values $0,1,2,3$ and are raised and lowered with the flat
Minkowski metric $\eta_{mn}={\rm diag}(-1,1,1,1)_{mn}$. Here, we view $e$ as the vierbein
for the underlying spacetime and $\omega$ as the spin
connection.
The Yang--Mills curvature $F$ then becomes
\bea
F = \left (
\begin{array}{cc}
R^{m}{}_n & T^m \\ 
0 & 0
\end{array}
\right ).
\eea
where $R = d \omega + \omega \wedge \omega$ is the Riemann curvature and
$T = d e + \omega \wedge e$ is the torsion of the connection.
We may work either with torsion-free $T = 0$ spacetimes or include it according to
the physics being probed. In the absence of torsion, the spin connection 
can be solved for as a function of the vierbein and the Yang--Mills equations
become the equations of harmonic curvature
\be
D^\mu R_{\mu\nu\rho\sigma}=2D_{[\rho} R_{\sigma]\nu}=0\, .
\ee
This requirement is weaker than Einstein's equations. Obvious solutions
are Ricci flat, Einstein and self-dual backgrounds so the model clearly has
a wide physical applicability.

Finally, now that the model couples to gravitational backgrounds, we obtain higher
spin fields by taking the vector field $V$ to be a tensor representation of the Poincar\'e
group. These can be decomposed in terms of tensor representations of the Lorentz
subgroup, so generically we find theories of higher spin fields 
$(f_\mu{}^{m_1\ldots m_s}, v_\mu{}^{m_1\ldots m_{s-1}}, \ldots , v_\mu )$.

\section{Minkowski Twisted Maxwell }

\label{Minkowski}

We make two simplifications. The background space is 
Minkowski ${\mathbb R}^{3,1}$ and the representation $R$ is 
the fundamental of the Yang--Mills Poincar\'e gauge group
$G = SO(3,1) \rtimes {\mathbb R}^4$.
In this case the Yang--Mills curvature vanishes so there no non-minimal coupling
in the detour operator ${\cal G}$. So we simply have what is known
as a twisted Maxwell  complex.
The fundamental representation acts naturally on a 5-vector of 1-forms,
\bea
V = \left ( \begin{array}{c} f^\rho \\ v \end{array} \right ) = 
\left ( \begin{array}{c} f_\mu{}^\rho \\ v_{\mu} \end{array} \right ) dx^\mu,
\eea
and we no longer distinguish between flat (Lie algebra) and curved 
(spacetime) indices using the latter in both cases.
Moreover, for a flat background the Riemann curvature, torsion, and spin connection
all vanish and the Yang-Mills potential is simply
\bea
A = \left (
\begin{array}{cc}
0 & \delta_\mu{}^\rho \\
0 & 0
\end{array}
\right )dx^\mu\, .
\eea
A simple computation yields the Lagrangian
\bea
L = - \quarter \left ( F_{\mu \nu} \right )^2 
       - \quarter \left (G_{\mu \nu} {}^{\rho} \right )^2, \label{CurvaturesS}
\eea
where the ``Maxwell''  curvatures 
(not to be confused with their background Yang--Mills counterpart in the previous Section)
\bea
F_{\mu \nu} &\equiv& \d_\mu v_\nu-\d_\nu v_\mu\ , \nn\\[4mm]
G_{\mu \nu} {}^{\rho} &\equiv&  \d_{\mu} f_{\nu} {}^{~\rho}
  -\d_{\nu} f_{\mu} {}^{~\rho}
  + \delta^{\rho} {}_{\mu}
  v_{\nu}
   -\delta^{\rho} {}_{\nu}
  v_{\mu}\, .
\eea

The gauge invariance $V\rightarrow V+ D\alpha$ becomes
\bea
f_{\mu}{}^\rho&\rightarrow& f_{\mu}{}^\rho+\partial_\mu \alpha^\rho +\delta_{\mu}^\rho\,\beta\, ,\nn\\[4mm]
v_\mu&\rightarrow& v_\mu\ +\ \partial_\mu \beta\, .
\eea
In 4 dimensions there are twenty fields $(f_\mu{}^\rho,v_\mu)$ and five gauge invariances
with parameters $(\alpha^\rho,\beta)$ 
so the model certainly describes a total of $20-2\times 5=10$ physical degrees of freedom.
(This is also obvious from the standpoint of five massless Yang--Mills vector matter fields.)
However, the partition of these modes into the irreducible Poincar\'e representations
of Wigner~\cite{Wigner:1939cj} is hardly clear from the Lagrangian~\eqn{CurvaturesS}.
To emphasize this point we expand this equation out as 
\bea
L &=& -\, \frac12 (\partial_\mu f_{\nu}{}^ \rho)^2
  +\frac12 (\d. f^{\rho})^2 
  -\frac12 (\partial_\mu v_\nu)^2
  +\frac12  (\d . v)^2  \nn \\[2mm]
  &-& m \, v^{\nu} (\d^{\rho} f_{\nu \rho}
  - \d_\nu f^{\rho}_{~\rho}) + (d-1)\, m^2 \, v . v\, .
  \eea
The top line is a sum of Maxwell actions but the second line includes cross terms and
an apparent mass term (here we have given the general result valid in $d$-dimensions).
We have included a mass parameter $m$ by na\"ive dimensional analysis. It can clearly take any
value we so choose and we will work in units $m=1$ for the remainder of the Article. 
It is important to note that this is a freedom peculiar to flat space. Upon considering more
general curved backgrounds, the parameter $m$ must be tuned to the gravitational coupling\footnote{This could be either a curse or blessing, see~\cite{Deser:2001dt} for a detailed analysis of this issue.}.

\section{Hamiltonian Helicity Analysis}

To determine the spectrum of the model we make a Hamiltonian analysis and
helicity decomposition. We treat the
time coordinate on a separate footing and denote spatial indices by $i,j,k,...=1,2,3$.
The following computation is completely standard (excellent references are~\cite{Arnowitt:1960es}),
but we sketch some details for completeness.

Firstly, introduce canonical momenta  $P^j$ and $\pi^j_{~\rho}$ by
\bea
P^j &= \ \frac{\d L}{\d \dot{v}_j}\
=&F_0^{~j}\ =\ \dot{v}^j - \d^j v_0 \label{pi^j}\, ,
\nn \\
\pi^j_{~\rho} &= \ \frac{\d L}{\d \dot{f}_j^{~\rho}}\
=&G_0^{~j} {}_{\rho}
= \dot{f}^j_{~\rho} - \d^j f_{0 \rho}
  - 2\, \delta_{\rho}^{~[0} v^{j]} \label{pi^jr}
\, .
\eea
Noting that the first order Lagrangian obtained by Legendre transformation
must take the form $L^{(1)}=P^jF_{0j}+\pi^j{}_\rho F_{0j}{}^\rho - \widehat H$
we rapidly find (suppressing spatial integrations $\int d^3x$)
\bea
L^{(1)}&=& P^j \dot{v}_j + \pi^j_{~\rho} \dot{f}_j^{~\rho} - H\, ,\nn\\[5mm]
H&=& \half \left [ (P_j)^2 + (\pi_j^{~\rho})^2 \right ]+ \quarter \left [ (F_{ij})^2 + (G_{ij} {}^{\rho})^2  \right ]
- \pi^j_{~0} v_j
\nn\\[2mm]
  &+&  v_0 \, \left [ \pi^j_{~j} - \d_j P^j  \right]
  - f_0^{~\rho}\ \d_j \pi^j_{~\rho} 
   \, .
\eea
Clearly, $v_0$ and $f_0^{~\rho}$ are Lagrange multipliers imposing
primary constraints
\bea
\pi^j_{~j} - \d_j P^j & = & 0\, , \nn \\[3mm]
\d_j \pi^j_{~\rho}     & = & 0\, . \label{eom2}
\eea

We now proceed by making a helicity decomposition, solving the constraints
and computing an action principle for physical degrees of freedom only.
Our helicity decomposition for general  1- and 2-index tensors is
\bea
Y_i    &=& Y^T _i + \d_i Y^L\, , \nn \\[3mm]
X_{ij} \! &=&  \!X^{Tt} _{ij} + 2 \d_{(i} X^T _{j)}
  + \half \Big(\delta_{ij} - \boxfrac{\d_i\d_j}{\Delta}\Big) X^S 
  + \boxfrac{\d_i\d_j}{\Delta} X^L
  + \epsilon_{ijk} \d^k X^A + 2 \d_{[i} X^{AT} _{j]}.\nn\\[1mm]
\eea
(where, for example, transverse objects are divergence free, so $\partial^i Y_i^T=0$).
We also heavily employ their inner products 
under a $\int d^3x$ integration
\bea
Y'{}_i Y^i&=&
Y'{}_i^{T} Y^{T  i}
-Y'^{L}\Delta Y^L\, ,
\nn\\[3mm]
X'_{ij}X^{ij}&=&
X'{}_{ij}^{Tt}X^{Tt ij}-2X'{}^{T}_j\Delta X^{T j}
+\half X'^S X^S+X'^L X^L 
\nn\\[1mm]&-&2 X'^A \Delta X^A -2 X'^{AT}_j\Delta X^{AT j}\, .\nn\\
\eea
Here the negative definite operator $\Delta=\partial_i\partial^i$ denotes the spatial Laplacian
which we take invertible. A useful mnemonic is that the number of
indices on fields now labels their helicity. Written out helicity by helicity the 
primary constraints~\eqn{eom2} are solved via
\be
\begin{array}{|c|c|}
\hline&\\[-3mm]
\mbox{Helicity}&\mbox{Constraints}\ 
\\[1mm]\hline&\\[-2mm]
\pm 1 &  \pi^{AT} _k  = -\pi^T _k
\\[2mm]\hline&\\[-3mm]
 & \pi^L = 0 \\[1mm]
0    & \pi^L _0  =  0 \\ [1mm]
    &P^L  =  \frac{1}{\Delta} \pi^S
\\[1mm]\hline
\end{array}
\label{constraints}
\ee
There are, of course, no constraints on the leading helicity~$\pm 2$ sector whose action reads
\bea
L^{(1)} _{\pm 2}& = & \pi^{Ttij} \dot{f}^{Tt}_{ij} 
   -  \left [ \half(\pi^{Ttij})^2
    + \half f^{Tt}_{ij} (-\Delta) f^{Ttij} \right ].
\eea
This consistently describes a physical massless spin two graviton.
The helicity zero sector is not much more difficult.
Upon substituting the constraints, $f_0^L$ decouples and making field redefinitions
\bea
q_0 &=& \sqrt{\frac{-2}{\Delta}} \pi^S\, , \nn\\
p_0     &=& \sqrt{\frac{-\Delta}{2}} \ (v^L - \half f^S)\, , \nn\\
\pi &=& \sqrt{\frac{-\Delta}{2}} \  \pi^A\, , \nn\\
\varphi &=& \sqrt{\frac{-\Delta}{2}} \ f^A\, ,
\eea
we find
\bea
L^{(1)}_0 &=& \pi \dot{\varphi}
  \ - \,  \half \left [ \pi^2
  + \varphi(-\Delta) \varphi \right ]  \nonumber \\[2mm]
 &  +& p_0\dot{q}_0  - \half \left [ p_0^2 
  + q_0 (-\Delta + 2\, ) q_0\right ] .
\eea
This describes a pair of physically consistent scalar fields, one
massless and one with mass $\sqrt{2}$. 
As we shall see in the following Section, the latter forms the zero helicity component 
of a physical massive vector field.

\section{Helicity 1 Hamiltonian Analysis}
\label{Danger}

The helicity~1 sector is more subtle. Although classically consistent, the model 
displays negative norm states when expanded about the trivial Lorentz invariant
field theoretic background. Firstly we perform the classical constraint analysis.

Imposing the helicity $\pm 1$ constraint as in~(\ref{constraints}), we find that
the combination $f^T_j + f^{AT} _j$ decouples and
\bea
L^{(1)}_{\pm 1} &=& \Pi^t \dot{\Phi} - H^{(1)}_{\pm 1}, \nn \\[3mm]
H^{(1)}_{\pm 1} &=& \half (\Pi^t \, \wt{M}\,  \Pi + \Pi^t \wt{N} \Phi
                                + \Phi \wt{P} \Phi)\, ,\label{action1}
\eea
where we have made field redefinitions 
packaged as a vector of $SO(2,1)$
\be
\Phi^T_j=\left(\begin{array}{c}
v_j^T\\[2mm]
f_{0j}^{\ T}\\[2mm]
\sqrt{-\Delta}\ (f_j^T-f^{AT}_j)\end{array}\right)\, ,\qquad \qquad
\Pi_j^T=
\left(\begin{array}{c}
P_j^T\\[2mm]
\pi_{0j}^{\ T}\\[2mm]
2\sqrt{-\Delta}\ \pi_j^T\end{array}\right)\, ,
\ee
and 
$$
\wt{M} =  
\left(
\begin{array}{ccc}
1&0&0\\[2mm]
0&-1&0\\[2mm]
0&0&1
\end{array}
\right)\, ,\qquad
\wt{N} =
 \left(
\begin{array}{ccc}
0&0&0\\[2mm]
-2&0&0\\[2mm]
0&0&0
\end{array}
\right)\, ,
$$
\vspace{.3cm}
\be
\wt{P} = \left(
\begin{array}{ccc}
-\Delta+2&\ \ 0\ \ &-\sqrt{-\Delta}\\[2mm]
0&\Delta&0\\[2mm]
-\sqrt{-\Delta}&0&-\Delta
\end{array}
\right)\, .
\ee
Throughout this and the following Sections
 we suppress the helicity~$\pm1$ labels~``$\ {}^T_j\ $''.
The dynamics are most easily analyzed via the second order 
form\footnote{An interesting rewriting of this action is in terms of an $SO(2,1)$ covariant
derivative $D=d+MN$, so that
$$
S_{\pm 1}^{(2)}=\frac12\   \frac{D\Phi^t}{dt} M  \frac{D\Phi}{dt} +\frac12\  \Phi^t (P+NMN)\Phi\, .
$$
The second term does, however, break the $SO(2,1)$ invariance.
} of the action~\eqn {action1}
\be
L^{(2)}_{\pm 1} = \half \dot{\Phi}^tM \dot{\Phi} + \dot{\Phi}^t N \Phi
  + \half \Phi P \Phi, \, 
\ee
where now
$$
M = \left (
  \begin{array}{ccc} 1 & 0 & 0 \\[2mm]
                     0 &-1 & 0 \\[2mm]
                     0 & 0 & 1
  \end{array} \right )\, ,\qquad \qquad
N = \left (
  \begin{array}{ccc} 0 & -\half & 0 \\[2mm]
                     \half & 0 & 0 \\[2mm]
                     0 & 0 & 0
  \end{array} \right )\, ,
 $$
\vspace{.3cm}
\be
P = \left (
  \begin{array}{ccc} \Delta - 3 & 0 & \sqrt{-\Delta} \\[2mm]
                     0 &-\Delta & 0 \\[2mm]
                     \sqrt{-\Delta} & 0 & \Delta
  \end{array} \right )\, .
\ee
The equations of motion 
\bea
\label{odeForm}
-M \ddot{\Phi} -2N \dot{\Phi} + P \Phi = 0.
\eea
are a second order matrix ODE.
Working in the eigenspace $\Delta=-k^2$
and considering wave solutions  $\Phi = \lambda e^{i \omega t}$,
then (\ref{odeForm}) becomes
\bea
\left ( M \omega^2 - 2i N \omega + P \right ) \lambda &=& 0\, .
\eea
The determinant of this matrix must vanish which yields
\bea
(k^2 -\omega^2 + 2)(k^2 -\omega^2)^2 &=& 0\, .
\eea
The zeros are precisely the relativistic dispersion 
relations of a single mass~$\sqrt{2}$ and two massless vector fields.
Observe that this mass eigenvalue agrees with that found in the
zero helicity sector so we obtain a pair of photons and a massive
vector. This is the spectrum quoted in the Introduction, we now 
analyze its quantization and stability.

\section{Quantization and Stability}
\label{quantization}

To quantize the model we expand the on-shell fields on plane wave solutions
\bea
\Phi &=& \sum_{i=1}^3 (f_i \alpha_i^\dagger + \overline{f_i} \alpha_i), \nn \\
\eea
where
\be
f_1 =\boxfrac54
  \left ( \begin{array}{c}
    0 \\[2mm] 1\\[2mm]  i
  \end{array} \right ) e^{ikt} \, ,\qquad
f_2 = f_1+ ik
  \left ( \begin{array}{c}
    1 \\[2mm]  \half t \\ [2mm]  \boxfrac i2 t
  \end{array} \right ) e^{ikt} \, ,
\ee
are photon solutions and
the massive vector solution is
\be
f_3 =
  \left ( \begin{array}{c}
    1 \\[2mm] -\half i \sqrt{k^2 + 2} \\[2mm] - \half k
  \end{array} \right ) e^{i\sqrt{k^2 +2}\  t} \, .
\ee
As we shall see, the massive vector subspace of the Hilbert space is perfectly physical
while the photon subspace is pathological. Already we see that the solution $f_2$ has
amplitude growing linearly in time. Mathematically this is a generalized eigenvector
solution to our system of PDEs.  Physically it can be interpreted in terms of 
a resonance between highly tuned wave solutions and indicates an
instability. Similar behavior has already been observed in the ghost
condensation mechanism of~\cite{Arkani-Hamed:2003uy} employed to obtain infra-red modifications of
Einstein gravity.

We now promote the Fourier coefficients  $(\alpha_i,\alpha^\dagger_i)$ to 
operators in a Fock space. Positivity of the classical energy and in turn stability can be studied through
the energy eigenvalues of single particle states. We will also analyze unitarity
of the model by computing norms of quantum states.

Imposing canonical equal time commutation relations of the fields and their 
momenta
\bea
[\Pi, \Phi^t] &=& -i \bf{1}\,  ,
\eea 
fixes the commutation relations of the creation and annihilation operators to
\be
\Omega\equiv [\alpha,\alpha^\dagger]=\left ( \begin{array}{cc|c}
    -\frac{2k}{25}  & -\frac{1}{5k} & 0 \\[1mm]
    -\frac{1}{5k} &0 & 0 \\[1mm]\hline
    0& 0 & \frac{1\phantom{^0}}{2 \sqrt{k^2 +2}}
  \end{array} \right )\, .
\ee
(the right hand side of this equation is the Wronskian of the solutions above).
As promised this is block diagonal and positive definite in the 
massive vector block. The zero on the diagonal already signals the
presence of zero norm states in the photonic Fock space.

The Hamiltonian may be expressed also in terms of Fock operators as
\be
H=\alpha^\dagger {\cal M}  \alpha\, ,\label{Qham}
\ee
with matrix
\be
{\cal M}=\left ( \begin{array}{cc|c}
    0& -5k^2 & 0 \\[2mm] -5k^2 & 2k^2(k^2-1)  & 0 \\[2mm]\hline 0 & 0 & 2(k^{2\phantom{^{0^0}}}\!\!\!\!\! + 2)
  \end{array} \right )
\, .
\ee
Taking into account the normalization of the symplectic form $\Omega$ we 
see that massive vectors states have both positive norms and energies 
with single particle, relativistic dispersion relation
\be
E=\sqrt{k^2+2}\, .\label{mass2}
\ee
The photonic Fock space is much more subtle.
Interestingly enough the eigenvalues of the matrix $\cal M$ can become negative
but are actually bounded below. However, consider a single particle state
\be
|1\rangle = \alpha^\dagger \lambda |0\rangle\, ,
\ee
where $|0\rangle$ is the Fock vacuum and $\lambda$ is some constant, complex
$3$-vector of coefficients.
Requiring $|1\rangle$ to be an energy eigenstate 
implies that
\be
H |1\rangle = \alpha^\dagger {\cal M} \alpha \ \alpha^\dagger \lambda |0\rangle
=\alpha^\dagger {\cal M} \Omega \lambda |0\rangle=E|1\rangle
\ee
and in turn the equality
\be
{\cal M} \Omega \lambda = E \lambda \, .
\ee
{\it I.e.}, we must diagonalize the effective Hamiltonian matrix
${\cal H}\equiv {\cal M}\Omega$
rather than simply ${\cal M}$. Explicitly
\be
{\cal H}=\left ( \begin{array}{cc|c}
    k & 0 & 0\\[2mm] \boxfrac{2k}{5} & k & 0 \\[1mm] \hline 0^{\phantom{0^{0^0}}} \!\!\!\!\!\!\!& 0 & \sqrt{k^2 + 2}
  \end{array} \right )\, . \label{effham}
\ee
Again we see that the massive vector decouples with dispersion relation~\eqn{mass2}.
While the only photon single particle energy eigenstate is \be|\gamma\rangle\equiv
a_2^\dagger |0\rangle\, ,\ee
with energy $E=k$ which is the correct Lorentz invariant dispersion relation for massless
excitations. The norm of this state $\langle \gamma|\gamma\rangle=0$, vanishes however.

We can also consider a general photonic single particle state 
$(\nu a_1^\dagger + \mu a_2^\dagger)|0\rangle$. Then denoting $\rho=\nu/\mu$
we find that states with $\rho$ inside the disc \be \Big |\rho+\frac{5}{2k^2}\Big|<\frac{5}{2k^2}\, ,
\label{bound}\ee
have positive norm, those on the boundary zero norm and those exterior to the disc negative norm
(the state $a^\dagger_1|0\rangle$ with 
$\rho=\infty$ also has negative norm).
The only single particle  state diagonalizing the Hamiltonian is the zero norm state
 $|\gamma\rangle$
corresponding to $\rho=0$.

Observe that positivity properties of norms are improved in the non-relativistic limit $k\rightarrow 0$,
for which any $\rho$ in the upper half plane solves~\eqn{bound}. Nonetheless even in this limit 
the non-unitarity difficulty persists. Another mechanism available to cure the instability
is to truncate the model by restricting physical states further to the cohomology of an appropriate
nilpotent operator. Explicitly, call the top 2$\times$2 block of the effective Hamiltonian 
in~\eqn{effham}~$\wh{\cal H}$. Then since any matrix obeys its own characteristic polynomial,
the matrix ${\cal N}\equiv \wh{\cal H}-k$ is nilpotent \be{\cal N}{\, }^2=0\, ,\ee and commutes with 
$\wh {\cal H}$.  The cohomology of ${\cal N}$ in the malevolent photonic single particle Fock space
is trivial, which is promising. We have not computed its cohomology for multiparticle 
states, but instead remark that this mechanism is unlikely to respect Lorentz invariance.

The presence of zero and negative norm states signals the breakdown of unitary evolution, as evidenced by the non-hermitean effective Hamiltonian matrix ${\cal H}$,  commensurate with
resonant classical single particle wave-functions growing linearly in time. Whether this instability
indicates the existence of other stable but possibly non-Lorentz invariant vacua,
or is  a runaway instability is an open problem deserving further study.
It seems likely that the addition of interparticle interactions is necessary to support a stable vacuum.

\section{Coherent State Evolution.}
\label{coherent_state}

Let us consider coherent states in the photonic Fock space\footnote{This analysis is similar in spirit 
to~\cite{Albrecht:1992kf}, where models with wrong sign potentials and squeezed states are analyzed.}
Denoting $\wh\alpha=(\alpha_1,\alpha_2)$ and similarly employing hats to denote
the top 2$\times$2  photonic block for matrices, 
coherent states diagonalizing the 
annihilation operators \be\wh\alpha|z\rangle=z|z\rangle\, ,\ee are simply
\be
|z\rangle=\exp(\wh\alpha^\dagger\wh\Omega^{-1}z)|0\rangle\, .
\ee
Here $z$ is a complex 2-vector and the coherent state associated
with the photon single particle state $|\gamma\rangle$ 
corresponds to $z_\gamma=\left(\begin{array}{c}0\\1\end{array}\right)$.

Its time evolution, given by\footnote{In quantum mechanics
coherent states evolve classically up to a phase corresponding to 
the zero point energy. As evidenced by~\eqn{Qham}, we have made the
usual field theoretic normal ordering renormalization so this factor is absent.}
\be
|z(t)\rangle=e^{iHt}|z\rangle\, ,
\ee
is easily computed to be
\be
z(t)=\left(\begin{array}{c} z_1\\[3mm] z_2+\frac{2ikt}{5} z_1 \end{array}\right)
e^{ikt}\, ,
\ee
which is the classical solution found above.
Therefore, as usual, coherent states are maximally classical.
The inner product for these states is 
\be
\langle w|z\rangle=\exp(w^\dagger \wh \Omega^{-1} z)\, .
\ee
Since $\wh \Omega$ is a real symmetric matrix, norms of photonic
coherent states
\be
\langle z|z\rangle =\exp(z^\dagger \wh \Omega^{-1} z)\, ,
\ee
are always positive. However, they are are not conserved in time
since evolution is no longer unitary (observe that the effective Hamiltonian ${\cal H}$
in~\eqn{effham} is not Hermitean). Instead we find that norms for the time evolved states
$|z(t)\rangle$ obey
\be
\langle z(t)|z(t)\rangle=\exp\Big(z^\dagger
\left(
\begin{array}
{cc}
\frac{8t^2k^5}{25} & -\frac{k(25+4ik^3t)}{5}\\[1mm]
-\frac{k(25-4ik^3t)}{5} & 2 k^3
\end{array}
\right)
z
\Big) \, .
\ee
Observe that the photon coherent state $|z_\gamma\rangle$ has a time independent
norm $\Big|\Big||z_\gamma\rangle\Big|\Big|^2=\exp(2k^3)$. In general, however, unitary evolution
is violated. In particular the state with $z=\left(\begin{array}{c}1\\0\end{array}\right)$, 
corresponding to $\rho=\infty$ in the notation of the previous Section, has norm behaving as
$\exp(4t^2k^5/25)$ for large times. This indicates that coherent combinations of the negative norm 
single particle states dominate the large time behavior of the model and are primarily 
responsible for its instability.

\section{Conclusions}
\label{Conclusions}

The Yang--Mills detour complex, obtained from an on-shell  Poincar\'e Yang--Mills twist of the
Maxwell complex along with a non-minimal coupling, yields a novel mechanism for coupling
higher spins to  gravitational backgrounds.
Even the simplest, flat, fundamental representation
version of the 
model, analyzed in depth here, has a rich spectrum though photon states have nonpositive norms.

There are many open questions and directions the model can taken
in. Firstly, vacua other than the usual Lorentz invariant background,
where all fields vanish, might be stable. Secondly, the
Yang--Mills gauge group~$G$ can be enlarged. Obvious generalizations are
to situations with conformal symmetry or supersymmetry where
$G$ can be the conformal or super Poincar\'e algebra~\cite{GSS}.

In general, given a complex, it often is possible to search for
projections to a smaller one where the projections and differentials
commute. ({\it I.e.}, one forms a commutative diagram.) Hence, one can
search for a smaller complex in which the zero norm and negative norm
states are excised~\cite{GSS}.

Another extremely interesting direction is to study models with
infinite towers of fields by taking Maxwell fields labeled by infinite
dimensional yet unitary representations of the Yang-Mills
algebra~${\mathfrak g}$. These present the possibility of a
fundamental theory with quantum consistency in the Lorentz invariant
vacuum. Moreover, one might even hope that genuine interparticle
interactions (rather than just ones to the background) would be
possible with the infinite number of fields as the loophole in the 
Coleman--Mandula theorem.

\section*{Acknowledgments}
It is a pleasure to thank Stanley Deser and Nemanja Kaloper for discussions and
encouragement. A.W. is indebted to the University of Auckland for its warm hospitality.
This work was supported by the National Science Foundation grant
PHY01-40365.

\end{document}